\documentclass[doublespacing]{elsart}

\usepackage{icarus}

\usepackage{natbib}

\usepackage{longtable}
\usepackage{lscape}
\usepackage{setspace}
\usepackage{lineno}
  
\bibpunct{(}{)}{;}{a}{,}{,}

\usepackage{graphicx}

\usepackage{amssymb}
\usepackage{wasysym}

\newcommand{\BibTeX}{ \textrm{B\kern-.05em\textsc{i\kern-.025em b}\kern-.08em
    T\kern-.1667em\lower.7ex\hbox{E}\kern-.125emX} }

\begin{document}

\begin{frontmatter}



\title{Rotational Characterization of Hayabusa II Target Asteroid (162173) 1999~JU3}


\author[mit,dtm]{Nicholas A. Moskovitz\thanksref{label}}
\author[ncu]{Shinsuke Abe}
\author[ncu]{Kang-Shian Pan}
\author[lco]{David J. Osip}
\author[mit]{Dimitra Pefkou}
\author[iafe]{Mario D. Melita}
\author[iafe]{Mauro Elias}
\author[ua]{Kohei Kitazato}
\author[ifa]{Schelte J. Bus}
\author[mit]{Francesca E. DeMeo}
\author[mit]{Richard P. Binzel}
\author[jsc]{Paul A. Abell}

\address[mit]{Massachusetts Institute of Technology, Department of Earth, Atmospheric and Planetary Sciences, 77 Massachusetts Avenue, Cambridge, MA 02139 (U.S.A.)}

\address[dtm]{Carnegie Institution of Washington, Department of Terrestrial Magnetism, 5241 Broad Branch Road, Washington, DC 20008 (U.S.A.)}

\address[ncu]{National Central University, Institute of Astronomy, 300 Jhongda Road, Jhongli, Taoyuan 32001 (Taiwan)}

\address[lco]{Carnegie Institution of Washington, Las Campanas Observatory, Colina El Pino, Casilla 601 La Serena (Chile)}

\address[iafe]{Universidad de Buenos Aires, Instituto de Astronomica y Fisica del Espacio, Buenos Aires (Argentina)}

\address[ua]{The University of Aizu, Research Center for Advanced Information Science and Technology, Ikki-machi, Aizu-Wakamatsu, Fukushima 965-8580 (Japan)}

\address[ifa]{University of Hawaii, Institute for Astronomy, 640 N. A'ohoku Place, Hilo, HI 96720 (U.S.A.)}

\address[jsc]{NASA Johnson Space Center, Astromaterials Research and Exploration Science Directorate, Houston, TX 77058 (U.S.A.)}

 \thanks[label]{Observations conducted while at Carnegie DTM, current address is MIT.}

\begin{center}
\scriptsize
Copyright \copyright\ 2012 Nicholas A. Moskovitz
\end{center}


%
%
%
%
%


\end{frontmatter}



\begin{flushleft}
\vspace{1cm}
Number of pages: \pageref{lastpage} \\
Number of tables: \ref{lasttable}\\
Number of figures: \ref{lastfig}\\
\end{flushleft}


\begin{pagetwo}{Rotational characterization of 1999 JU3}

Nicholas A. Moskovitz\\
Department of Earth, Atmospheric and Planetary Sciences\\
Massachusetts Institute of Technology\\
77 Massachusetts Avenue\\
Cambridge, MA 02139, USA. \\
\\
Email: nmosko@mit.edu\\
Phone: (617) 253-1268 \\

\end{pagetwo}

 \linenumbers

\begin{abstract}

The Japanese Space Agency's Hayabusa II mission is scheduled to rendezvous with and return a sample from the near-Earth asteroid (162173) 1999 JU3. Previous visible-wavelength spectra of this object show significant variability across multiple epochs which could be the result of a compositionally heterogeneous surface. We present new visible and near-infrared spectra to demonstrate that thermally altered carbonaceous chondrites are plausible compositional analogs, however this is  a tentative association due to a lack of any prominent absorption features in our data. We have also conducted a series of high signal-to-noise visible-wavelength observations to investigate the reported surface heterogeneity. Our time series of visible spectra do not show evidence for variability at a precision level of a few percent. This result suggests two most likely possibilities. One, that the surface of 1999 JU3 is homogenous and that unaccounted for systematic effects are causing spectral variation across epochs. Or two, that the surface of 1999 JU3 is regionally heterogenous, in which case existing shape models suggest that any heterogeneity must be limited to terrains smaller than approximately 5\% of the total surface area. These new observations represent the last opportunity before both the launch and return of the Hayabusa II spacecraft to perform ground-based characterization of this asteroid. Ultimately, these predictions for composition and surface properties will be tested upon completion of the mission.

\end{abstract}




\section{Introduction \label{sec.intro}}

In June of 2010 the Japanese Space Agency's Hayabusa mission successfully completed the first ever sample return from an asteroid \citep[][and references therein]{fujiwara06,abe06b,nakamura11}. The success of this mission has spawned a successor, Hayabusa II, which is scheduled to rendezvous with and return a sample from the asteroid (162173) 1999 JU3. This object was discovered by the Lincoln Near-Earth Asteroid Research (LINEAR) survey on 1999 May 10. It is an Apollo-class near-Earth asteroid with one of the lowest values of $\Delta v$ for any object with an absolute magnitude less than H=20\footnote{http://echo.jpl.nasa.gov/$\sim$lance/delta\_v/delta\_v.rendezvous.html}, and is thus a highly favorable space mission target. The current Hayabusa II timeline proposes a launch in 2014 or 2015, rendezvous with the asteroid in 2018, and return of a sample to Earth in 2020. Like its predecessor this mission provides the opportunity to directly test the validity of ground-based analytical techniques with pre-encounter predictions for the physical and chemical properties of the asteroid \citep[e.g.][]{binzel01b}.

Previous observations of 1999 JU3 were conducted following its discovery in 1999 and during its favorable apparition in 2007. Ground-based photometry determined a light curve period of about 7.63 hours \citep{abe08,muller11}, which has since been confirmed with photometry from the 2012 apparition \citep{kim12}. A weighted average of space and ground-based observations in the mid-infrared suggest an effective diameter of 870 $\pm$ 30 meters and an albedo of 6.954 $\pm$ 0.005\% \citep{hasegawa08,campins09,muller11}. Prior to the present study, visible-wavelength spectra had been obtained in May 1999, July 2007, and September 2007 (Figure \ref{fig.spec}, Table \ref{tab.spec}). These data display marked variability in spectral slope and in the presence of absorption features. The spectrum from 1999 shows UV absorption short-ward of about 0.65 $\mu$m and a featureless neutral slope at longer wavelengths \citep{binzel01}. The July 2007 data, which suffer from the lowest signal-to-noise, show an overall red slope with a pronounced absorption feature $\sim10\%$ deep centered between 0.6 and 0.7 $\mu$m \citep{vilas08}. The September 2007 spectrum is a combination of data from two nights (Table \ref{tab.spec}) and is largely featureless with neutral slope \citep{vilas08}. Though these three spectra are broadly consistent with taxonomic classification in the C-complex, such pronounced variability, in this case at the level of $\sim10$\%, is highly unusual. 

This spectral variability has been attributed to a compositionally heterogenous surface \citep{vilas08}. The absorption feature between 0.6 and 0.7 $\mu$m in the July 2007 spectrum was linked to Fe$^{2+} \rightarrow$ Fe$^{3+}$ charge transfer transitions in iron-bearing phyllosilicates. The absence of this feature in the May 1999 and September 2007 spectra suggests a non-uniform distribution of these aqueous alteration byproducts. The possible existence of hydrated minerals on the surface of 1999 JU3 is of considerable interest for the purposes of mission planning and the potential sampling of primitive Solar System material. Such surface heterogeneity could be inherited from the asteroid's parent body or could be exogenous in origin, as is believed to be the case for the "black boulder" seen by  Hayabusa I on Itokawa \citep{fujiwara06} and for the dark material imaged by the Dawn spacecraft on 4 Vesta \citep{reddy12a}.

We present observations of 1999 JU3 from the 2012 apparition, the last favorable observing window before both the scheduled launch and return of the Hayabusa II spacecraft. Visible and near-infrared spectra were obtained to constrain the composition of the asteroid and to investigate claims of a non-uniform surface. Other investigators obtained visible spectra during the 2012 observing window \citep{sugita12,vilas12,lazzaro13}. These other data are not directly addressed here, though we note that none of these works found evidence for a 0.7 $\mu$m absorption feature or any repeatable spectral variability through multiple rotation periods. We also present here new visible-wavelength photometry to constrain the rotational phases of the asteroid accessed by the 2012 spectroscopic observations. Our observations and reduction techniques are presented in Section \ref{sec.obs}. Analysis of these data are presented in Section \ref{sec.analysis}. We summarize and discuss the results in Section \ref{sec.disc}.

\section{Observations and Data Reduction \label{sec.obs}}

Data presented here were taken with seven different instruments at six telescopic facilities and include visible and near-infrared spectroscopy, and visible-wavelength photometry. With the exception of one near-infrared spectrum from 2007, all data were obtained between October of 2011 and July of 2012. Tables \ref{tab.spec} and \ref{tab.phot} summarize the observing circumstances.

\subsection{Visible-Wavelength Spectroscopy: LDSS3 \label{subsec.vis}}

New visible wavelength spectra were obtained with LDSS3 at the Magellan Clay 6.5m telescope on the nights of UT 2012 June 1-3, coinciding with the opposition of the asteroid at a phase angle of 0.2$^\circ$ on UT June 1 at 07:30. The instrument was operated with a 1.5"-wide and 4'-long slit and its VPH-All grism. These settings produced a single-order spectrum with a useful spectral range of 0.44-0.94 $\mu m$ at an average resolution of about 430 and a dispersion of 1.89 \AA/pixel. Each individual spectrum represents a set of 4 x 180s exposures, split between two different nod positions approximately 20" apart along the slit.

Solar analog stars were observed several times throughout each night for calibration purposes (Table \ref{tab.spec}). With the exception of HD149182, the calibration stars were selected from a list of well-vetted analogs that were used extensively throughout the SMASS survey \citep{bus02}. HD149182 was also included because of its close proximity ($<10^\circ$) to the asteroid and was selected based on its photometric colors and G2V spectral classification. The high degree of consistency amongst our measured spectra (Section \ref{subsec.variability}) confirms that HD149182 is a good solar analog.

Reduction of these spectra employed standard IDL and IRAF routines, including a set of IRAF tools developed specifically for LDSS3. Combined He, Ne and Ar arc lamp spectra were obtained at the beginning and end of each night to provide dispersion solutions. Exposures of a white screen illuminated by a quartz lamp were used for flat field calibration.

\begin{figure}[]
\begin{center}
\includegraphics[width=14cm]{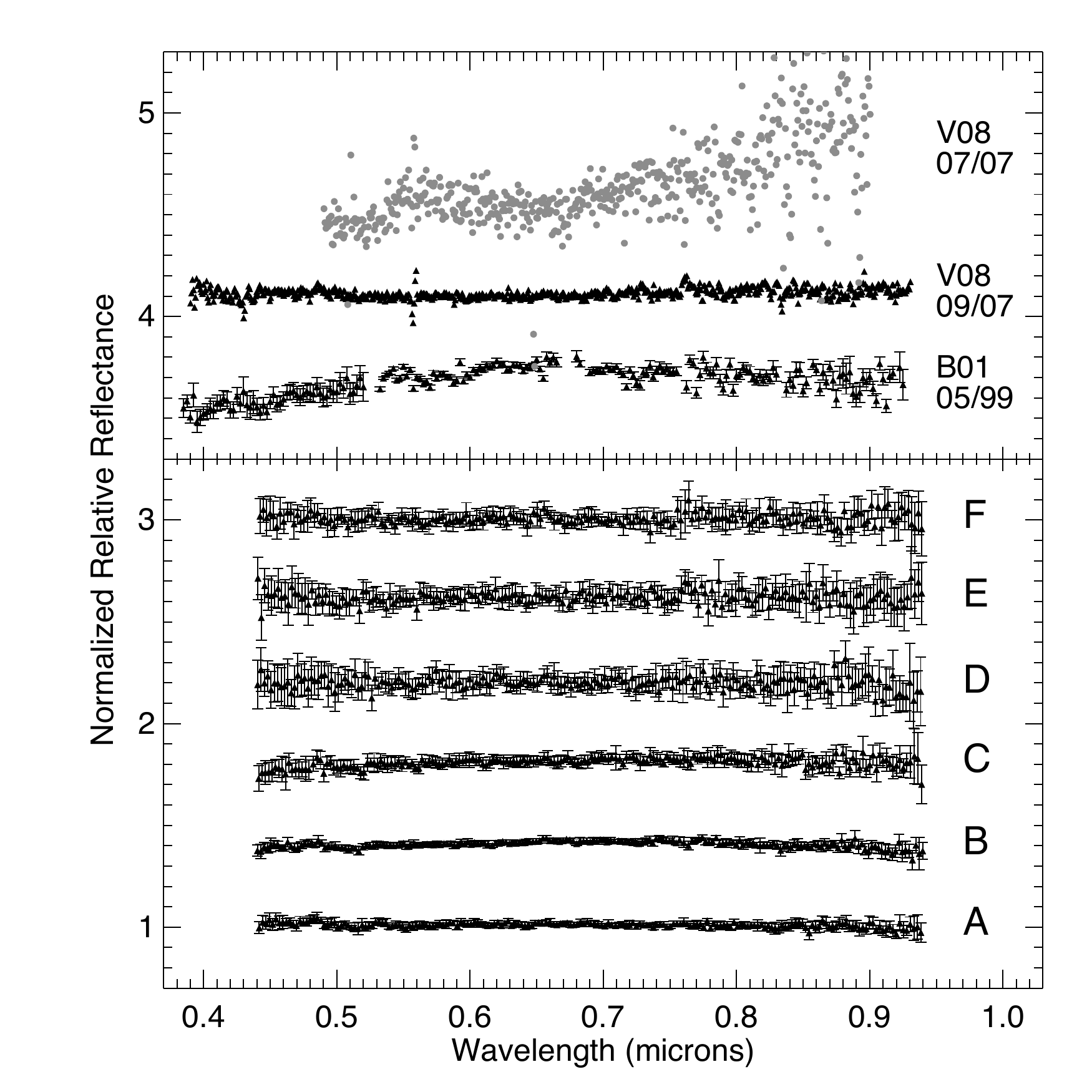}
\end{center}
\caption[]{Normalized visible-wavelength reflectance spectra of 1999 JU3. The data are vertically offset for clarity. The top panel shows the three previous visible spectra with the month/year of observation and respective references B01 = \citet{binzel01} and V08 = \citet{vilas08}. The July 2007 data is shown in grey to distinguish it from the adjacent spectrum. Our spectra are in the bottom panel and are chronologically ordered from bottom to top with spectra A and B taken on June 1, C on June 2, and D-F taken on June 3. The midpoint times for each of these observations are given in Table \ref{tab.subobs}. The progressive decrease in signal-to-noise in spectra A-F is due to the opposition effect, increasing geocentric and heliocentric distances, and increasing background from a waxing moon.
} 
\label{fig.spec}
\label{lastfig}
\end{figure}

The extracted asteroid spectra were divided by each of the solar analogs (typically $\sim10$ individual observations of solar analogs throughout the night). The best solar analog for a given asteroid spectrum was chosen based on minimization of residual telluric features long-wards of $\sim0.7~\mu$m. After division by an analog, individual points in the spectra were then rejected based on a 0.25 $\mu$m boxcar sigma-clipping routine with a prescribed cutoff of 3 sigma. In all cases less than 10\% of the data points were rejected and those that were removed were predominantly distributed at the extremes of the spectra outside the useful spectral range of 0.44-0.94 $\mu m$. Generally, less than 10 pixels were removed from within this final trimmed spectral range. After cleaning, the asteroid spectra were then re-binned by a factor of 13 pixels to produce a dispersion interval very close to 0.0025 $\mu$m per channel, similar to that of the 1999 spectrum \citep{binzel01} and a factor approximately 3 times than that of the 2007 data \citep{vilas08}. The error bar on each binned dispersion element was set to the standard deviation of the original 13 pixels. The data were finally normalized at 0.55 $\mu$m to produce relative reflectance spectra (Figure \ref{fig.spec}). 

The wavelength-averaged signal-to-noise ratio (S/N) of these spectra vary from $\sim60$ on the first night (A, B in Figure \ref{fig.spec}), to $\sim30$ on the second night (C), to $\sim20$ on the third night (D, E, F). Since exposure times were held constant, this  drop in S/N is likely caused by several effects. First, spectra A and B were taken at solar phase angles less then $1^\circ$ (Table \ref{tab.spec}) and thus may have captured an opposition surge in brightness. Second, the moon was both waxing and increasingly close to the target throughout the run (Table \ref{tab.spec}), thus causing a progressive increase in background flux. Third, the target's geocentric and heliocentric distances were both increasing, causing a drop in brightness across the three nights.

\subsection{Near-Infrared Spectroscopy: FIRE and SpeX \label{subsec.nir}}

Near-infrared spectra were obtained in June and July of 2012 with FIRE at the Magellan Baade 6.5m telescope and in September 2007 with SpeX \citep{rayner03} at NASA's Infrared Telescope Facility (IRTF; Table \ref{tab.spec}). FIRE was operated in its high-throughput prism mode with a 0.8 x 50" slit. These settings produced single-order spectra at a resolution of approximately 400 from 0.8 to 2.45 $\mu$m. Individual exposures were limited to 180s to avoid saturation of telluric emission features and of thermal emission from the instrument and telescope at wavelengths longer than about 2.2 $\mu$m. Exposures were obtained in ABBA sequences with the object offset along the slit by 9" for each nod position. On UT 2012 June 5 a total of 4 x 180s exposures were obtained. On UT 2012 July 10, when the asteroid was nearly 2 magnitudes fainter, a total of 18 x 180s exposures were obtained. 

At the IRTF, SpeX was configured in its low resolution (R = 250) prism mode with a 0.8" slit for wavelength coverage from 0.8 to 2.5 $\mu$m. All observations were made with the telescope operating in a standard ABBA nod pattern with individual exposures of 120s. On UT 2007 September 18 a total of 40 individual exposures were obtained, on UT 2007 September 20 a total of 50 individual exposures were obtained.

For both near-infrared spectrographs the slit mask was oriented along the parallactic angle at the start of each observation to minimize the effects of atmospheric dispersion. As with the visible-wavelength spectra, solar analogs were observed to correct for telluric absorption and to remove the solar spectrum from the measured reflectance (Table \ref{tab.spec}). Reduction of the SpeX data followed \citet{sunshine04}, reduction of the FIRE data employed an IDL package designed for the instrument and based on the Spextool pipeline \citep{cushing04}.

A composite visible/near-infrared spectrum of 1999 JU3 is shown in Figure \ref{fig.full}. The asteroid is a C-type in the \citet{demeo09} taxonomic system. The visible portion is a weighted average of the six LDSS3 spectra in Figure \ref{fig.spec}. The near-infrared portion is a weighted average of the SpeX and FIRE data, and thus represents a combination of data across multiple epochs. Other than some variability in slope, no significant differences were seen amongst the near-infrared spectra within the S/N of the data.

\subsection{Visible-Wavelength Photometry: IMACS, Lulin, Tenagra II, and Bosque Alegre \label{subsec.phot}}

Photometric observations (Table \ref{tab.phot}) were conducted with IMACS at the Magellan Baade 6.5m telescope at Las Campanas Observatory in Chile, with a Princeton Instruments EEV 1k CCD at the Lulin 1m telescope in Taiwan, with a SITe 1k CCD at the Tenagra II 0.81m telescope in Arizona, and with a Tektronics 1k CCD at the Bosque Alegre 1.5m telescope in Argentina. These data were taken on ten separate nights between October 2011 and June 2012.

IMACS was operated using the $f$/2 camera, which has a 27.5' field-of-view covered by a mosaic of eight 2k x 4k CCDs with plate scales of 0.2 "/pixel. We employed only one of the eight chips with a subraster about 400x400" in size. We used a Bessell R-band filter with exposure times of 60 seconds. These observations spanned approximately 3 hours per night over three consecutive nights. The Princeton Instruments CCD at Lulin Observatory has a roughly 11' field of view with a plate scale of 0.516 "/pixel. These observations were conducted with a Johnson-Cousins R-band filter over the course of approximately 2 hours on a single night. Exposure times were set to 300 seconds. The SITe 1k CDD at the Tenegra II telescope has a 14.8' field-of-view with a plate scale of 0.87 "/pixel. A Johnson-Cousins R-band filter was used. These observations spanned approximately 4.5 hours per night over 4 non-consecutive nights with individual exposure times set to 360 seconds. The Tektronics 1k at the Bosque Alegre 1.5m has a plate scale of 0.33 "/pixel and a field-of-view of 5.7'. Observations were conducted with a Johnson R-band filter, spanning approximately 3.5 hours on two consecutive nights. Exposure times were 240 seconds the first night and 200 seconds the second.

Reduction of these data employed standard IRAF procedures. Aperture photometry was performed with aperture radii roughly 4 times the measured seeing and a background annulus equal to twice that separation. Some individual exposures were rejected based on contamination from nearby field stars. Photometric calibration was achieved for the IMACS data with observations of the RU 149 Landolt standard field \citep{landolt92}. All other photometry was performed differentially relative to at least one half dozen on-chip field stars and then scaled to match the calibrated IMACS data. A full table of the collected photometry is included in the online Supplementary Data.

\section{Rotational and Compositional Analyses \label{sec.analysis}}

Our visible-wavelength photometry is used to construct a phase-folded rotational light curve of 1999 JU3. This provides the means to constrain the specific rotational phases accessed by the spectroscopic observations and to quantitatively address the previously reported surface heterogeneity. The combined visible/near-IR spectrum is used to place preliminary constraints on composition.

\subsection{Rotational Light Curve \label{subsec.lightcurve}}

\citet{muller11} find a rotation period for 1999 JU3 of 7.63 $\pm$ 0.01 hours and \citet{kim12} find a period of 7.625 $\pm$ 0.003 hours. We use these periods as a guide to phase-fold our broadband photometry (Figure \ref{fig.lc}). The highest S/N data came from the three nights of observations at Magellan (Table \ref{tab.phot}). The magnitudes of all other observations are scaled to the mean value of the photometrically calibrated Magellan data. As with previous light curve measurements \citep{muller11,kim12}, our data suggest a small peak-to-peak amplitude of about 0.2 magnitudes. Attempts have been made to exclude data points contaminated by background field stars, however the low galactic latitude of the asteroid ($<20^\circ$ for all observations) certainly adds to the scatter in the measured photometry.

The light curve in Figure \ref{fig.lc} has been phase folded with a synodic period of 7.631 hours relative to the first IMACS observation on UT 2012-04-05 at 06:40 (JD 2456022.778). Light time corrections have been applied relative to this first IMACS observation. This light curve is wrapped such that the first and last few points are repeated at phases $<0$ and $>1$. Robustly fitting a light curve period is difficult because the photometric errors are comparable to the amplitude of the light curve and our best quality photometry from IMACS does not span a full rotation period. A unique period is not found when employing the Fourier analysis of \citet{polishook12}. Thus we attempt to manually constrain the period based on three features in the IMACS data: a subtle peak at phase = 0.2, a minimum at phase = 0.4 and another peak around phase = 0.6. The period is adjusted in increments of 0.001 hours until these three features are roughly aligned for all data. We find that these features are not aligned for periods within $\pm$0.005 hours of 7.631 hours. This result is consistent with the range of periods suggested by \citet{muller11} and \citet{kim12}, but is not meant to be a more accurate determination.

\subsection{Non-detection of Spectroscopic Variability \label{subsec.variability}}

\begin{figure}[]
\begin{center}
\includegraphics[width=14cm]{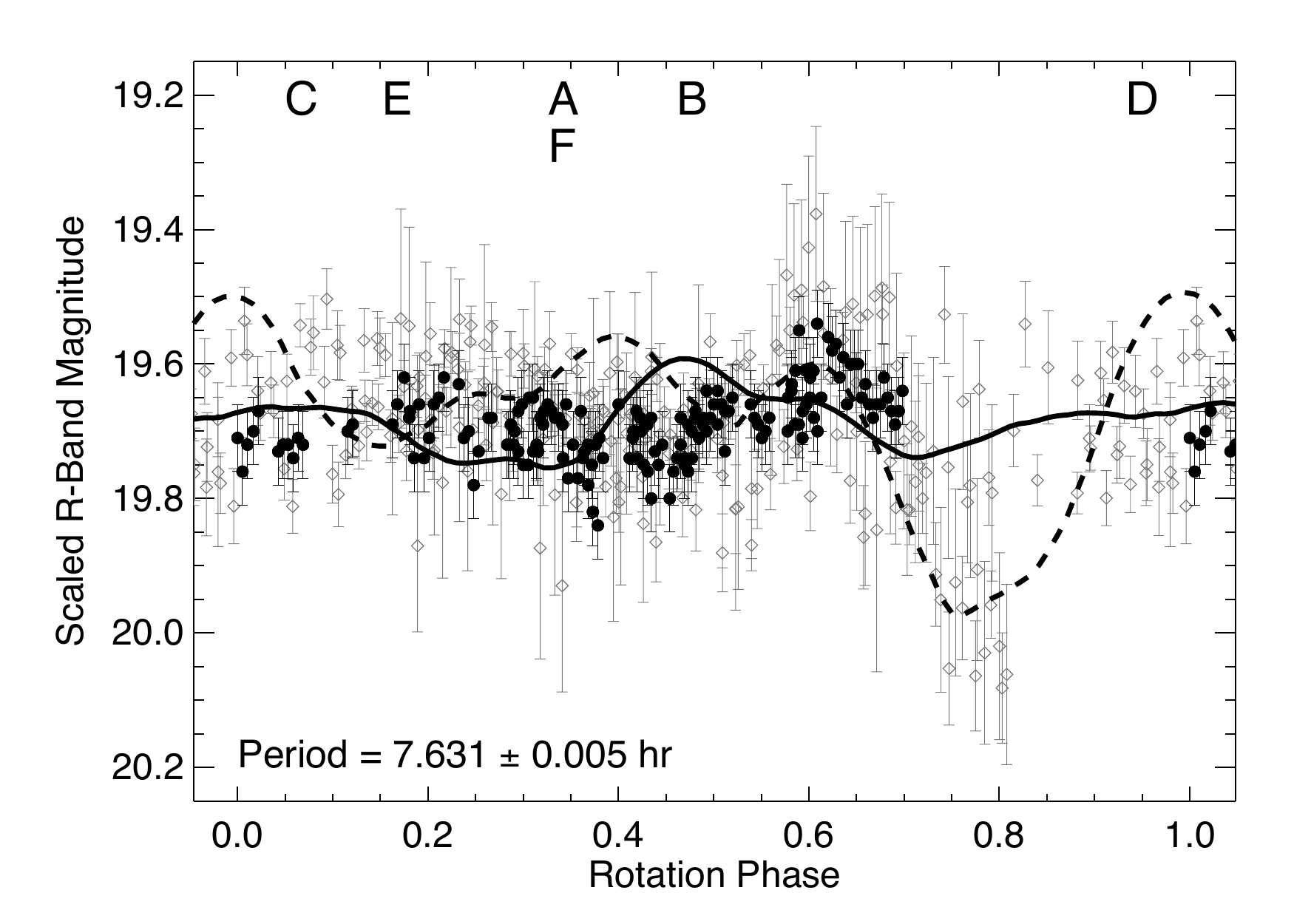}
\end{center}
\caption[]{Rotational light curve of 1999 JU3. Filled circles represent the IMACS Magellan data. Open diamonds represent our additional observations (Table \ref{tab.phot}) with magnitudes scaled to the IMACS data. The solid line represents the predicted light curve from the \citet{muller11} model, the dashed is the model prediction from \citet{kawakami10}. The data have been phase folded to a synodic period of 7.631 hours. A rotation phase of zero corresponds to the start of IMACS observations on UT 2012-04-05 at 06:40 (JD 2456022.778). The phases accessed by the visible-wavelength spectroscopic observations are indicated by the letters A-F. The total time for each spectrum (4 x 180 seconds plus readout and telescope offsets) is comparable to the width of the letters. Spectroscopic coverage was obtained over approximately 60\% of rotation phases at a longitudinal resolution of $\sim45^\circ$.
} 
\label{fig.lc}
\label{lastfig}
\end{figure}

The phase folded light curve is used to provide rotational context for each of the new spectroscopic observations (Figure \ref{fig.lc}). This phase folding demonstrates that rotational coverage was obtained over approximately 60\% of the asteroid's surface at a longitudinal sampling of $\sim45^\circ$. The spectroscopic observations from previous epochs \citep{binzel01,vilas08} are not included in Figure \ref{fig.lc} because current best estimates for the rotation period \citep[e.g.][]{kim12} still have an associated error that is large enough to preclude linking observations separated by several years.

\begin{figure}[]
\begin{center}
\includegraphics[width=14cm]{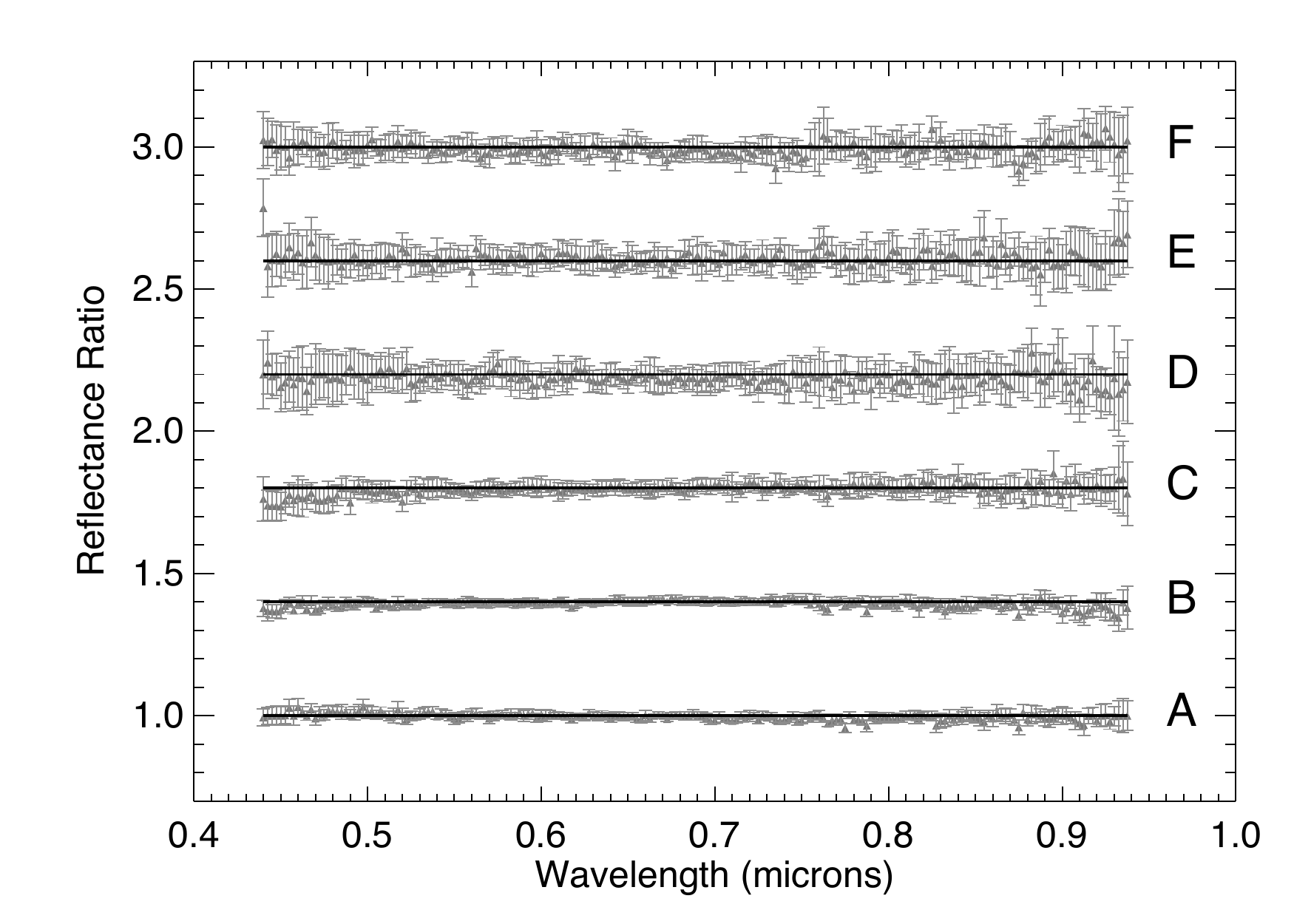}
\end{center}
\caption[]{Visible-wavelength reflectance ratios of 1993 JU3 relative to a median combination of all spectra. These reflectance ratios have been offset by multiples of 0.4 units. Deviations from a slope-zero line (black curves) would indicate spectroscopic heterogeneity. No variability is detected within the S/N of the data, though slight ($<4\%$) deviations from a flat line are seen at the shortest wavelengths in spectra B and C.
} 
\label{fig.ratios}
\label{lastfig}
\end{figure}

We compute reflectance ratios to investigate the possibility of a heterogenous surface (Figure \ref{fig.ratios}). These ratios are computed by median combining spectra A through F and then dividing that median into each of the individual spectra. Any deviations from a flat line would indicate the presence of a heterogenous surface. Typical deviations are around 1\%, with the largest (just under 4\%) seen at the bluest wavelengths in spectrum C. However, all deviations fall within the S/N of the data and thus suggest that the regions on the surface accessed by our observations are homogenous to within a few percent.

Previously reported spectroscopic variability around 0.8 $\mu$m \citep{lazzaro13} is not seen in our data set. This non-detection suggests the possibility raised by these authors, that it may be attributed to either instrumental optical fringing or residual effects from the telluric water absorption band centered near 0.8 $\mu$m.

The rotational context for these spectroscopic data provides a test of our observation and reduction techniques. Figure \ref{fig.lc} shows that spectra A and F were obtained at nearly identical rotation phases. The fact that these spectra are indistinguishable (Figure \ref{fig.ratios}), but taken on different nights and calibrated with different solar analogs (Table \ref{tab.spec}), is a nice validation of the consistency of our methodology.

\subsection{Spectroscopic Surface Coverage \label{subsec.surface}}

Shape models of 1999 JU3 based on photometry from 2007 have been independently developed by \citet{kawakami10} and \citet{muller11}. In both cases, the small amplitude of the light curve produces nearly spherical shapes. These models are based upon different data sets, but because of issues of low S/N, do not offer a single unique solution. Unsurprisingly then, neither shape model produces light curves that match our data well (Figure \ref{fig.lc}). Clearly the development of a new shape model that fits photometry from 2007 and 2012 is an important task for future work. Due to this ambiguity it is difficult to absolutely state the surface latitudes and longitudes accessed by the full ensemble of spectroscopic observations dating back to 1999. However we can compare model predictions for the spectroscopically accessed surface regions to illustrate the plausible spatial extent of any non-uniform surface features.

The spectra of 1999 JU3 from September 2007 are fully consistent with the new data presented here (Figure \ref{fig.spec}). Any surface heterogeneity must then be restricted to regions accessed by the observations in May 1999, when the asteroid displayed UV absorption short-ward of about 0.65 $\mu$m \citep{binzel01} and in July 2007 when the asteroid showed a prominent absorption feature between 0.6 and 0.7 $\mu$m  \citep{vilas08}. These regions would have to be limited in extent to avoid violating the lack of absorption features in the other spectra.

Table \ref{tab.subobs} presents the two sets of predicted sub-observer latitudes and longitudes for all visible-wavelength spectroscopic observations. Figure \ref{fig.map} illustrates the sub-observer points predicted by the two shape models. Though these sub-observer points are not meant to be absolute and may change with future improvements to the shape model, their relative separations are informative. In particular, we are most interested in the distance that separates the sub-observer points for featureless spectra (September 2007, June 2012) from the sub-observer points for spectra that display absorption bands (May 1999, July 2007; B and V in Figure \ref{fig.map}). We can roughly calculate the physical distance between sub-observer points by assuming a spherical body with a diameter of 870m, a reasonable approximation for this asteroid \citep{kawakami10,muller11}. The physical distance between sub-observer points is then just the distance along the great circle that connects these points.

According to the \citet{kawakami10} model, the observations from May 1999 were taken at a sub-observer point about 180m away from the next closest point which corresponds to spectrum F (June 2012). And the observations from July 2007 correspond to a sub-observer point about 160m away from the September 2007 spectra. According to the \citet{muller11} model, the observations from May 1999 were taken about 140m from the September 2007 points, and the observations from July 2007 correspond to a sub-observer point about 200m away from spectrum D (June 2012). Taken as a whole, and independent of which shape model is employed, these calculations suggest that the reported spectroscopic variability would be a consequence of regional surface heterogeneity on spatial scales smaller than about 400m in extent (200m in radius). Regions of this size would occupy about 5\% of the asteroid's total surface area.

The ambiguity in shape models emphasizes that rotational information is not correlated over multiple years. This is indicated by the horizontal bands in Figure \ref{fig.map}. The sub-observer points for those spectra with absorption features could lie anywhere along these bands. This provides some freedom for the spectrally anomalous regions  to ``hide" in the gaps between those sub-observer points corresponding to featureless spectra. However, the sub-observer points for both the \citet{binzel01} and \citet{vilas08} spectra would have to be shifted by very specific amounts to reconcile the full suite of visible spectra. This issue is further complicated upon consideration of the sub-observer hemispherical footprints of our spectroscopic observations (see online Supplementary Figure A). These footprints indicate that we accessed a significant majority of the surface, leaving very little surface area in which spectroscopically anomalous regions could exist.

\subsection{Compositional Analogs \label{subsec.composition}}

\begin{figure}[]
\begin{center}
\includegraphics[width=14cm]{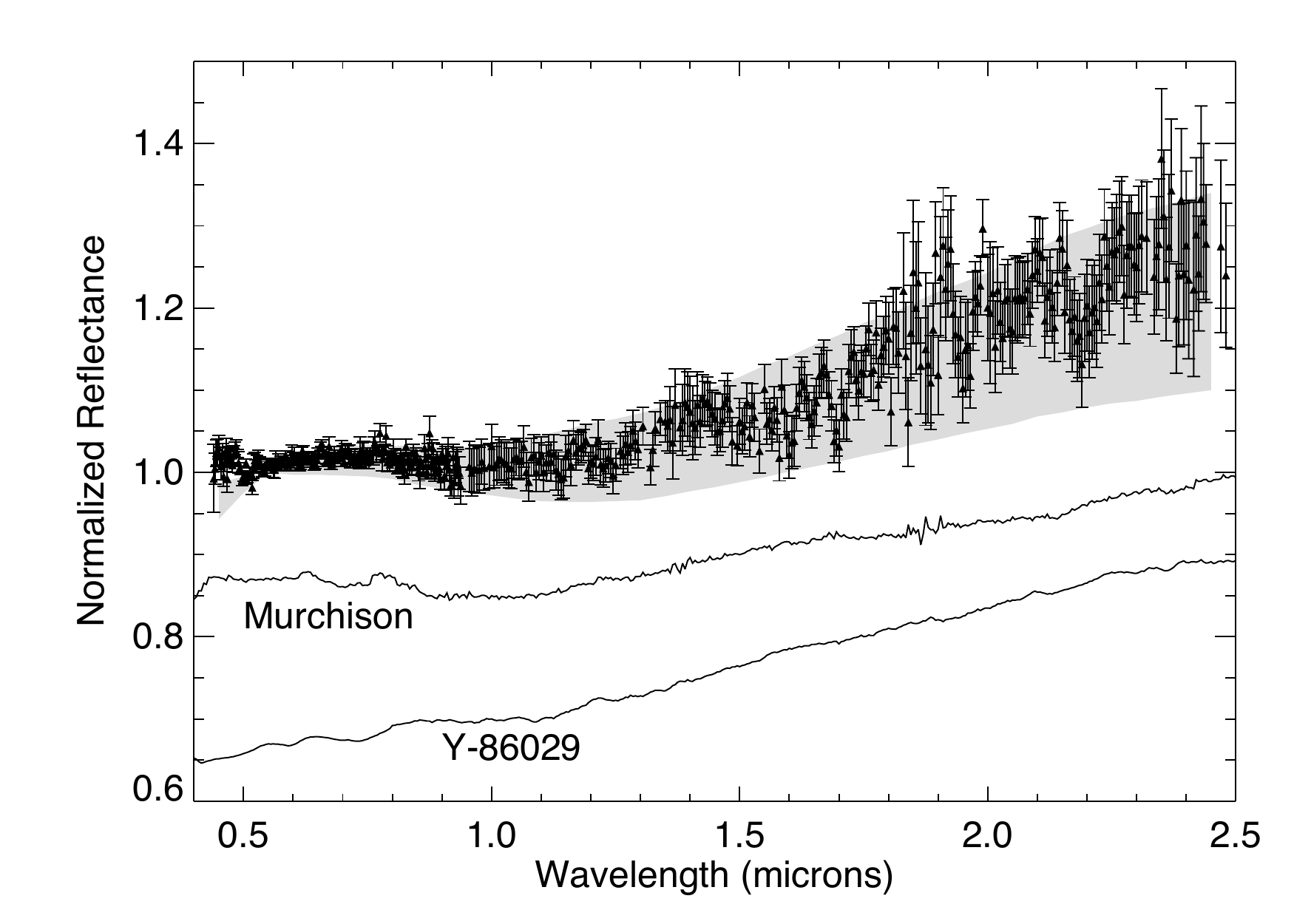}
\end{center}
\caption[]{Composite visible (LDSS3) and near-infrared (SpeX and FIRE) spectrum of 1999 JU3. The grey region in the background represents the envelope for C-type spectra in the \citet{demeo09} taxonomic system. In spite of the low signal-to-noise in the near-infrared and the presence of residual telluric features around 1.4 and 1.9 $\mu$m, this spectrum shows no evidence for prominent absorption bands, though subtle features may be present around 0.9 $\mu$m and short-ward of 0.7 $\mu$m. The two best spectral matches from our $\chi^2$ search through RELAB are a thermally altered sample of Murchison and the unusual CI chondrite Yamato 86029. These meteorite spectra have been normalized at 1 $\mu$m and then offset by -0.15 and -0.3 units respectively.
} 
\label{fig.full}
\label{lastfig}
\end{figure}

With the full visible/near-infrared spectrum (Figure \ref{fig.full}) we can make preliminary statements about the composition of 1999 JU3. However, these statements must be interpreted with some skepticism due to the lack of prominent absorption features and issues of degenerate compositional interpretations when dealing with such featureless spectra \citep[e.g.][]{vernazza09,reddy12b,reddy12c}. A preliminary attempt at compositional analysis is justified by the potential for predictions to be directly tested with a returned sample. We base this analysis solely on our new observations; see \citet{vilas08} for an interpretation of previous visible-wavelength spectra.

Our visible and near-infrared spectra confirm a C-type taxonomic classification, but show no evidence for the previously detected 0.7 micron phyllosilicate band \citep{vilas08} or a prominent UV drop-off short-wards of about 0.65 $\mu$m \citep{binzel01}. The data may indicate weak bands around 0.9 $\mu$m and short-ward of 0.7 $\mu$m, but these features are near the limit of the S/N and are not robust detections. Without any strong features to directly analyze we employ spectral matching techniques to constrain plausible compositional analogs. Comparison spectra were obtained from a July 2012 release of Brown University's RELAB database \citep{pieters83} containing reflectance spectra for just over 17,000 samples. Our search began by selecting only those samples with a spectral range overlapping that of the asteroid and only those samples with an albedo within 5\% of that measured for 1999 JU3. This narrowed the list of comparison spectra to 8024 samples. For each of these, either the asteroid or the meteorite data were interpolated to match the wavelengths of the spectrum with the smallest dispersion interval. Both spectra were then normalized at the minimum asteroid wavelength (0.44 $\mu$m) before a reduced $\chi^2$ difference was calculated. 

The two meteorite spectra with the lowest $\chi^2$ values are shown in Figure \ref{fig.full}. These meteorites are a thermally altered sample of the CM carbonaceous chondrite Murchison and the unusual CI chondrite Yamato 86029. The reduced $\chi^2$ values for these meteorites were 2.89 and 3.25 respectively. These were clearly the best fits as the next lowest $\chi^2$ was 6.75 for a sample of a mid-ocean ridge basalt (RELAB ID RB-CMP-023) that was ground to particle sizes 125-250 $\mu$m. This spectrum of oceanic basalt matched the general slope of 1999 JU3 but showed a subtle 1 $\mu$m absorption band and thus is unlikely to be a good analog.

The mean albedos of the two best-fit samples (6\% for Murchison, 3\% for Yamato) are by design of the search algorithm close to the $\sim$7\% measured for 1999 JU3 \citep{hasegawa08,campins09,muller11}. The two best-fit spectra both correspond to the smallest available particle size sorting, $<63~\mu$m for Murchison and $<125~\mu$m for Yamato. The Murchison spectrum represents one in series of heating experiments where the temperature of the sample was raised in increments of 100$^\circ$C from 300-1000$^\circ$C \citep{hiroi96}. At each temperature increment a reflectance spectrum was measured. Heating was shown to reduce spectral slope and decrease the depth of absorption bands. The Murchison spectrum corresponding to a temperature of 900$^\circ$C was the closest match to 1999 JU3. Yamato 86029 has unusual textural properties and a chemical composition indicative of a rare class of CI chondrites that experienced moderate thermal alteration at temperatures of 500-600$^\circ$C \citep{tonui03}. 1999 JU3 has been previously linked to carbonaceous chondrites \citep{vilas08,lazzaro13}, the analysis here expands upon these results by specifically identifying thermally altered samples as the best available spectral analogs.

\section{Discussion \label{sec.disc}}

We have presented new observations of the Hayabusa II target asteroid 1999 JU3 taken during the favorable 2012 apparition, the last opportunity before both the launch and return of the spacecraft to perform ground-based characterization. Visible wavelength photometry was used to produce a rotational light curve. A period of 7.631 $\pm$ 0.005 hours and a small amplitude of a few tenths of a magnitude are consistent with independent light curve measurements \citep{muller11,kim12}. Visible and near-infrared spectra were used to constrain taxonomy and composition. The asteroid is a C-type in the \citet{demeo09} taxonomic system and the best spectral analogs are thermally altered carbonaceous chondrites. However, the lack of any prominent absorption features makes this compositional association tentative. High quality spectra that extend into the near-UV (0.3-0.4 $\mu$m) or the 3 $\mu$m region could provide further compositional clues if diagnostic absorption bands are detected \citep[e.g.][]{hiroi93,rivkin02} and may serve as a test of the collisional and thermal evolution of primitive asteroid types \citep{vilas96}.

The putative link to carbonaceous chondrites and particularly those that are thermally altered is interesting and merits some preliminary speculation. CM and CI chondrites are primitive meteorites with appreciable quantities of organics and water of hydration \citep{cloutis11a,cloutis11b}. The possible return of such samples would be relevant to issues of astrobiology and to understanding primordial chemistry in the solar nebula \citep[e.g.][]{morby00}. The suggested link to heated carbonaceous chondrites could imply that surface temperatures on 1999 JU3 were significantly higher at some point in its past. There is roughly a 50\% probability that the chaotic orbital evolution of 1999 JU3 resulted in a perihelion inside of 0.5 AU (its current perihelion distance in 0.96 AU), which translates to surface temperatures of approximately 500 K \citep{marchi09,michel10}. This is comparable to the in-situ alteration temperatures experienced by Yamato 86029, but is less than the 900$^\circ$C experienced by the Murchison sample. Nevertheless, this highlights that thermally altered primitive meteorites could be relevant to understanding the composition of 1999 JU3. Ultimately, the results from Hayabusa II will provide the first-ever direct test of such compositional interpretations from largely featureless asteroid spectra.

We have also presented a time series of visible spectra to address previously reported spectroscopic variability \citep{vilas08,lazzaro13}. Our spectra show no variability at the level of a few percent. There are several possible explanations for this non-detection. First, the asteroid's surface may be spectroscopically and compositionally uniform, in which case some unaccounted for systematic effects would be the cause of spectral variation across epochs. The hemispherical footprints of our spectral observations cover nearly the entire surface of the asteroid (Supplementary Figure A), thus supporting this hypothesis. Variations in spectral slope are common with asteroid observations and can be attributed to observational issues such as weather, phase angle \citep{sanchez12,reddy12d}, or the use of imperfect solar analogs, but it is unclear how these issues would result in variable absorption features. Other than a moderately high airmass of $\sim1.9$ for the \citet{binzel01} spectrum, we find no obvious observational reasons why the two spectra with absorption features should be different. 

Multi-epoch spectral variability could alternatively be linked to recent alteration events that induced global or near global changes in surface properties. Planetary encounters \citep{binzel10} or impacts \citep{richardson04} can alter reflectance properties, however these seem unlikely because two such events would be required in just 13 years to explain the observed spectroscopic variability from 1999 to 2007 to 2012. 

\begin{figure}[]
\begin{center}
\includegraphics[width=14cm]{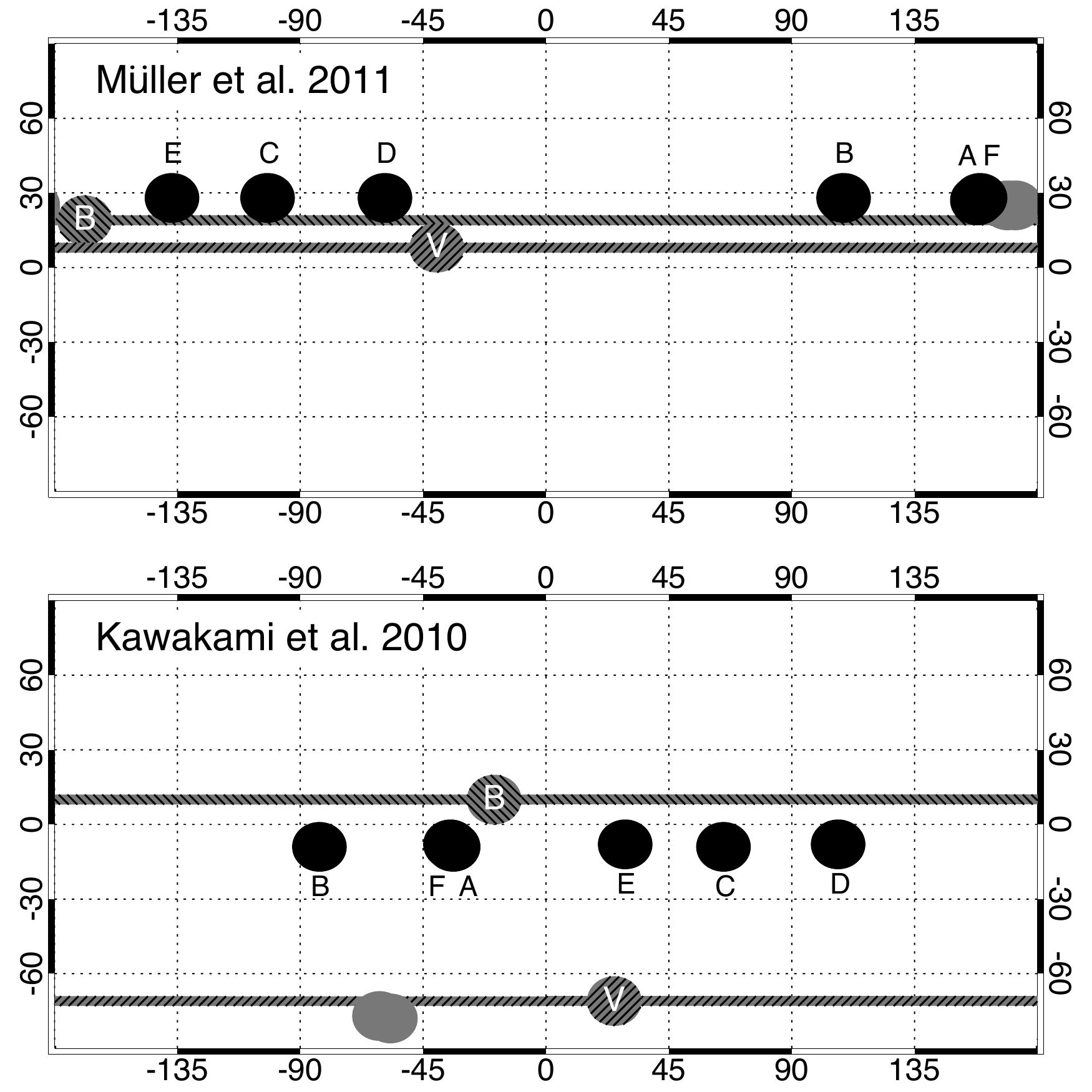}
\end{center}
\caption[]{Cylindrical projections of 1999 JU3's surface with circles indicating the sub-observer latitudes and longitudes from the two available shape models \citep{kawakami10,muller11}. The circles are not meant to indicate the physical extent of the surface accessed by the observations. Black circles represent sub-observer points from our new 2012 observations and are labeled A-F. Grey circles represent sub-observer points for the observations from 1999 and 2007. The two spectra that display absorption features are indicated with hatched regions and labeled B \citep{binzel01} and V \citep{vilas08}. Surface regions that produce absorption features must be limited in extent to avoid overlapping those corresponding to featureless spectra. The loss of rotational information across epochs (see text) suggests that the locations of these regions relative to the 2012 points could lie anywhere along the horizontal bands. This demands that any heterogeneity on the surface be confined to regions no more than a few hundred meters in extent.
} 
\label{fig.map}
\label{lastfig}
\end{figure}

A final explanation is that the surface of 1999 JU3 is in fact heterogenous. In this case spectral variability could result from the obscuration of surface regions by local topography as a consequence of changing viewing aspect. However, the orbital longitudes of both the Earth and the asteroid were very similar during the May 1999 and June 2012 observations. This would then require a highly irregular shape (i.e. large topographic relief) to result in significant obscuration and spectral variability. A highly irregular shape is inconsistent with current shape models \citep{kawakami10,muller11}. Therefore, spectral variability caused by a non-uniform surface would have to be attributed to rotational effects. The combination of rotational data and time resolved spectra provides a means to address this possibility. Existing shape models \citep{kawakami10,muller11} suggest that the maximum physical separation of sub-observer points for the full ensemble of spectroscopic data is about 200 meters (assuming that the asteroid is 870 meters in diameter; Figure \ref{fig.map}). Therefore any surface heterogeneity must be confined to regions no larger than about 400 meters in extent, which corresponds to a size independent surface area fraction of around 5\%. However, we note again that current shape models do not predict light curves consistent with photometry across multiple epochs. Therefore, the analysis here constrains the {\it plausible} spatial extent of non-uniform surface features, improved shape models will be required to refine the spectroscopically accessed terrains indicated in Figure \ref{fig.map}. 

Few asteroids, particularly sub-km objects like 1999 JU3, have been so extensively observed across multiple epochs. However, non-uniform surfaces are generally believed to be rare. The long-time exception is asteroid 4 Vesta, one of very few objects where variability over multiple rotation periods has been confirmed with remote observations. Vesta displays regional and hemispherical color heterogeneity, in some cases well in excess of several 10's of percent \citep[e.g.][]{bobrovnikoff29,gaffey97,reddy12a}. However, Vesta is much larger than 1999 JU3, is a different spectral class, and thus may not be the best object for comparison.

There have been reports of rotational spectroscopic heterogeneity amongst C-class asteroids. For instance, 10 Hygiea, 105 Artemis, 135 Hertha, and 776 Berbericia each show variability at levels less than about 5\% in the depths of their 0.7 $\mu$m features \citep[][]{rivkin02,vilas08}. Detecting such subtle variability is intrinsically difficult and as far as we are aware remains to be verified for any asteroid over multiple rotation periods. Asteroid 21 Lutetia, which was visited by the Rosetta spacecraft, shows variability in visible-wavelength spectral slope at the level of a few percent \citep{barucci12}. The S/N of our data for 1999 JU3 may be insufficient to detect such muted variability. But again, these examples of spectral variability occur on asteroids that are much larger than 1999 JU3.

The former target of the European Space Agency's Marco Polo mission (175706) 1996 FG3 and asteroid 25143 Itokawa, the target of the Hayabusa I mission, are examples of well-studied asteroids in a size regime closer to that of 1999 JU3. 1996 FG3 shows prominent ($\sim10$'s of percent) spectral variability in the near-IR \citep{deleon11}. The origin of this variability remains a mystery. Itokawa displays up to 15\% color variability on its surface \citep{saito06,abe06a}. This non-uniformity is limited to regions on the order of 10's to 100 meters in extent or at most a few percent of Itokawa's surface area. This extent is similar to the constraints we have placed on 1999 JU3's surface, though it remains unclear whether such small regions could significantly influence ground-based observations that only capture hemispherical averages. Ground-based observations of Itokawa prior to arrival of Hayabusa I did not exhibit any variability \citep{binzel01b,abell07}. 

For now, the issue of heterogeneity on the surface of 1999 JU3 remains an open question that may find some resolution from updated shape models. The arrival of the Hayabusa II spacecraft will ultimately determine whether the full ensemble of visible spectra are truly indicative of a non-uniform surface. More generally, the possibility of multiple planetary missions to near-Earth asteroids in the coming decades will provide an opportunity to further investigate the extent of surface heterogeneity on sub-km asteroidal bodies, and thus provide clues about formational and evolutionary processes in the Solar System.

\ack

We are grateful to Vishnu Reddy and Faith Vilas for their insightful reviews. We thank David Polishook for his helpful input on the lightcurve analysis. This work includes data obtained at the Magellan 6.5m telescopes located at Las Campanas Observatory in Chile and at NASA's IRTF located on Mauna Kea in Hawaii, which is operated by the University of Hawaii under Cooperative Agreement no. NNX-08AE38A with the National Aeronautics and Space Administration, Science Mission Directorate, Planetary Astronomy Program. Support for this project was provided to N.M. by the Carnegie Institution of Washington Department of Terrestrial Magnetism, the National Aeronautics and Space Administration through the NASA Astrobiology Institute (NAI) under Cooperative Agreement No. NNA04CC09A, and through the National Science Foundation Astronomy and Astrophysics Postdoctoral Fellowship.

\label{lastpage}



\begin{thebibliography}{}

\bibitem[M. Abe et al.(2006)]{abe06a} Abe, M. and 12 co-authors, 2006. Near-infrared spectral results of asteroid Itokawa from the Hayabusa spacecraft. Science 312, 1334-1338.

\bibitem[S. Abe et al.(2006)]{abe06b} Abe, S. and 16 co-authors, 2006. Mass and Local Topography Measurements of Itokawa by Hayabusa. Science 312, 1344-1347.

\bibitem[Abe et al.(2008)]{abe08} Abe, M. and 24 co-authors, 2008. Ground-based observational campaign for asteroid 162173 1999 JU3. Lunar Planet. Sci. 39. Abstract 1594.

\bibitem[Abell et al.(2007)]{abell07} Abell, P.~A., Vilas, F., Jarvis, K.~S., Gaffey, M.~J., and Kelley, M.~S., 2007. Mineralogical composition of (25143) Itokawa 1998 SF36 from visible and near-infrared reflectance spectroscopy: Evidence for partial melting. Met. Planet. Sci. 42, 2165-2177.

\bibitem[Barucci et al.(2012)]{barucci12} Barucci, M.~A. and 12 co-authors., 2012. Overview of Lutetia's surface composition. Planet. Space. Sci. 66, 23-30.

\bibitem[Binzel et al.(2001a)]{binzel01} Binzel, R.~P., Harris, A.~W., Bus, S.~J., and Burbine, T.~H., 2001a. Spectral Properties of Near-Earth Objects: Palomar and IRTF Results for 48 Objects Including Spacecraft Targets (9969) Braille and (10302) 1989 ML. Icarus 151, 139-149.

\bibitem[Binzel et al.(2001b)]{binzel01b} Binzel, R.~P., Rivkin, A.~S., Bus, S.~J., Sunshine, J.~M., and Burbine, T.~H., 2001b. MUSES-C target asteroid (25143) 1998 SF36: A reddended ordinary chondrite. Meteorit. Planet. Sci. 36, 1167-1172.

\bibitem[Binzel et al.(2010)]{binzel10} Binzel, R.~P., Morbidelli, A., Merouane, S., DeMeo, F.~E., Birlan, M., Vernazza, P., Thomas, C.~A., Rivkin, A.~S., Bus, S.~J., and Tokunaga, A.~T. Earth encounters as the origin of fresh surfaces on near-Earth asteroids. Nature 463, 331-334.

\bibitem[Bobrovnikoff(1929)]{bobrovnikoff29} Bobrovnikoff, N.~T., 1929. The spectra of minor planets. Lick Obs. Bull. 14 (No. 407), 18-27.

\bibitem[Bus and Binzel(2002)]{bus02} Bus, S.~J. and Binzel, R.~P., 2002. Phase II of the Small Main-Belt Asteroid Spectroscopic Survey: A Feature-Based Taxonomy. Icarus 158, 146-177.

\bibitem[Campins et al.(2009)]{campins09} Campins, H., Emery, J.~P., Kelley, M., Fern\'andez, Y., Licandro, J., Delb\'o, M., Barucci, A., and Dotto, E., 2009. Spitzer observations of spacecraft target 162173 (1999 JU3). A\&A 503, L17-L20.

\bibitem[Cloutis et al.(2011a)]{cloutis11a} Cloutis, E.~A., Hiroi, T., Gaffey, M.~J., Alexander, C.~M.~O'D., and Mann, P., 2011a. Spectral reflectance properties of carbonaceous chondrites: 1. CI chondrites. Icarus 212, 180-209.

\bibitem[Cloutis et al.(2011b)]{cloutis11b} Cloutis, E.~A., Hudon, P., Hiroi, T., Gaffey, M.~J., and Mann, P., 2011b. Spectral reflectance properties of carbonaceous chondrites: 1. CM chondrites. Icarus 216, 309-346.

\bibitem[Cushing et al.(2004)]{cushing04} Cushing, M.~C., Vacca, W.~D. and Rayner, J.~T., 2004. Spextool: A Spectral Extraction Package for SpeX, a 0.8Ð5.5 Micron Cross-Dispersed Spectrograph. PASP 116, 363-376.

\bibitem[de Le\'on et al.(2011)]{deleon11} de Le\'on, J., Moth\'e-Diniz, T., Licandro, J., Pinilla-Alonso, N., and Campins, H., 2011. New observations of asteroid (175706) 1996 FG3, primary target of the ESA Marco Polo-R mission. A\&A 530, L12, 1-4.

\bibitem[DeMeo et al.(2009)]{demeo09} DeMeo, F.~E., Binzel, R.~P., Slivan, S.~M., and Bus, S.~J., 2009. An extension of the Bus asteroid taxonomy into the near-infrared. Icarus 202, 160-180.

\bibitem[Fujiwara et al.(2006)]{fujiwara06} Fujiwara, A. and 21 co-authors, 2006. The rubble-pile asteroid Itokawa as observed by Hayabusa. Science 312, 1330-1334.

\bibitem[Gaffey(1997)]{gaffey97} Gaffey, M.~J., 1997. Surface lithologic heterogeneity of asteroid 4 Vesta. Icarus 127, 130-157.

\bibitem[Hasegawa et al.(2008)]{hasegawa08} Hasegawa, S. and 9 co-authors, 2008. Albedo, Size, and Surface Characteristics of Hayabusa-2 Sample-Return Target 162173 1999 JU3 from AKARI and Subaru Observations. Publ. Astron. Soc. Japan 60, S399-S405.

\bibitem[Hiroi et al.(1996)]{hiroi96} Hiroi, T., Zolensky, M.~E., Pieters, C.~M., and Lipschutz, M.~E., 1996. Thermal metamorphism of the C, G, B, and F asteroids seen from the 0.7 $\mu$m, 3 $\mu$m and UV absorption strengths in comparison with carbonaceous chondrites. Met. Planet. Sci. 31, 321-327.

\bibitem[Hiroi et al.(1993)]{hiroi93} Hiroi, T., Pieters, C.~M., Zolensky, M.~E., and Lipschutz, M.~E., 1993. Evidence of Thermal Metamorphism on the C, G, B, and F Asteroids. Science 261, 1016-1018.

\bibitem[Kawakami et al.(2010)]{kawakami10} Kawakami, K., Abe, M., Hasegawa, S., and Kasuga, T., 2010. Observation of the post Hayabusa mission target asteroid (162173) 1999 JU3. Japanese Society of Planetary Sciences 19, 4-11.

\bibitem[Kim et al.(2012)]{kim12} Kim, M.-J. and 9 co-authors, 2013. Optical observations of NEA 162173 (1999 JU3) during the 2011-2012 apparition. A\&A, accepted for publication.

\bibitem[Landolt(1992)]{landolt92} Landolt, A.~U., 1992. UBVRI photometric standard stars in the magnitude range 11.5 $<V<$ 16.0 around the celestial equator. AJ 104, 340-371.

\bibitem[Lazzaro et al.(2013)]{lazzaro13} Lazzaro, D., Barucci, M.~A., Perna, D., Jasmim, F.~L., Yoshikawa, M., and Carvano, J.~M.~F., 2013. Rotational spectra of (162173) 1999 JU3, the target of the Hayabusa 2 mission. A\&A 549, L2, 1-4.

\bibitem[Marchi et al.(2009)]{marchi09} Marchi, S., Delb\'o, M., Morbidelli, A., Paolicchi, P., and Lazzarin, M., 2009. Heating of near-Earth objects and meteoroids due to close approaches to the Sun. Mon. Not. R. Astron. Soc. 400, 147-153.

\bibitem[Michel and Delb\'o(2010)]{michel10} Michel, P. and Delb\'o, M., 2010. Orbital and thermal evolutions of four potential targets for a sample return space mission to a primitive near-Earth asteroid. Icarus 209, 520-534.

\bibitem[Morbidelli et al.(2000)]{morby00} Morbidelli, A., Chambers, J., Lunine, J.~I., Petit, J.~M., Robert, F., Valsecchi, G.~B., and Cyr, K.~E., 2000. Source regions and timescales for the delivery of water to the Earth. Met. Planet. Sci. 35, 1309-1320.

\bibitem[M\"uller et al.(2011)]{muller11} M\"uller, T.~G. and 17 co-authors, 2011. Thermo-physical properties of 162173 (1999 JU3), a potential flyby and rendezvous target for interplanetary missions. A\&A 525, A145, 1-6.

\bibitem[Nakamura et al.(2011)]{nakamura11} Nakamura, T. and 21 co-authors, 2011. Itokawa Dust Particles: A Direct Link Between S-Type Asteroids and Ordinary Chondrites. Science 333, 1113-1116.

\bibitem[Pieters(1983)]{pieters83} Pieters, C.~M., 1983. Strength of Mineral Absorption Featuresin the Transmitted Component of Near-Infrared Reflected Light: First Results From RELAB. J. Geophys. Res. 88, 9534-9544.

\bibitem[Polishook et al.(2012)]{polishook12} Polishook, D. and 19 co-authors, 2012. Asteroid rotation periods from the Palomar Transient Factory survey. Mon. Not. R. Astron. Soc. 421, 2094-2108.

\bibitem[Rayner et al.(2003)]{rayner03} Rayner, J., Toomey, D.~W., Onaka, P.~M., Denault, A.~J., Stahlberger, W.~E., Vacca, W.~D., and Cushing, M.~C., 2003. SpeX: A Medium-Resolution 0.8Ð5.5 Micron Spectrograph and Imager for the NASA Infrared Telescope Facility. PASP 115, 362-382.

\bibitem[Reddy et al.(2012a)]{reddy12a} Reddy, V. and 24 co-authors, 2012a. Color and albedo heterogeneity of Vesta from Dawn. Science 336, 700-704.

\bibitem[Reddy et al.(2012b)]{reddy12b} Reddy, V., Gaffey, M.~J., Abell, P.~A., and Hardersen, P.~S., 2012b. Constraining albedo, diameter and composition of near-Earth asteroids via near-IR spectroscopy. Icarus 219, 382-392.

\bibitem[Reddy et al.(2012c)]{reddy12c} Reddy, V., Le Corre, L., Hicks, M., Lawrence, K., Buratti, B., Abell, P., Gaffey, M.~J., and Hardersen, P.~S., 2012c. Composition of near-Earth asteroid 2008 EV5: Potential target for robotic and human exploration. Icarus 221, 678-681.

\bibitem[Reddy et al.(2012d)]{reddy12d} Reddy, V. and 11 co-authors, 2012d. Photometric, spectral phase and temperature effects on Vesta and HED meteorites: Implications for Dawn mission. Icarus 217, 153-168.

\bibitem[Richardson et al.(2004)]{richardson04} Richardson, J.~E., Melosh, H.~J. and Greenberg, R., 2004. Impact-Induced Seismic Activity on Asteroid 433 Eros: A Surface Modification Process. Science 306, 1526-1529.

\bibitem[Rivkin et al.(2002)]{rivkin02} Rivkin, A.~S., Howell, E.~S., Vilas, F., and Lebofksy, L.~A., 2002. Hydrated minerals on asteroids: The astronomical record.  In: Bottke, W.~F., Cellino, A., Paolicchi, P., Binzel, R.~P. (Eds.), Asteroids III. Univ. Arizona Press, Tucson, pp. 235-253.

\bibitem[Saito et al.(2006)]{saito06} Saito, J. and 33 co-authors, 2006. Detailed images of asteroid 25143 Itokawa from Hayabusa. Science 312, 1341-1344.

\bibitem[Sanchez et al.(2012)]{sanchez12} Sanchez, J.~A., Reddy, V., Nathues, A., Cloutis, E.~A., Mann, P., and Hiesinger, H., 2012. Phase reddening on near-Earth asteroids: Implications for mineralogical analysis, space weathering and taxonomic classiciation. Icarus 220, 36-50.

\bibitem[Sugita et al.(2012)]{sugita12} Sugita, S., Kuroda, D., Kameda, S., Hasegawa, S., Kamata, S., Abe, M., Ishiguro, M., Takato, N., and Yoshikawa, M., 2012. Visible spectroscopic observation of asteroid 162173 (1999 JU3) with the Gemini-S telescope. American Astronomical Society, 44th DPS meeting, Abstract 102.02.

\bibitem[Sunshine et al.(2004)]{sunshine04} Sunshine, J.~M., Bus, S.~J., McCoy, T.~J., Burbine, T.~H., Corrigan, C.~M., and Binzel, R.~P., 2004. High-calcium pyroxene as an indicator of igneous differentiation in asteroids and meteorites. Meteorit. Planet. Sci. 39, 1343-1357.

\bibitem[Tonui et al.(2003)]{tonui03} Tonui, E.~K., Zolensky, M.~E., Lipschutz, M.~E., Wang, M.-S., and Nakamura, T., 2003. Yamato 86029: Aqueously altered and thermally metamorphosed CI-like chondrite with unusual textures. Met. Planet. Sci. 38, 269-292.

\bibitem[Vernazza et al.(2009)]{vernazza09} Vernazza, P., Brunetto, R., Binzel, R.~P., Perron, C., Fulvio, D., Strazzulla, G., and Fulchignoni, M., 2009. Plausible parent bodies for enstatite chondrites and mesosiderites: Implications for LutetiaÕs fly-by. Icarus 202, 477-486.

\bibitem[Vilas and Sykes(1996)]{vilas96} Vilas, F. and Sykes, M., 1996. Are low-albedo asteroids thermally metamorphosed?. Icarus 124, 483-489.

\bibitem[Vilas(2008)]{vilas08} Vilas, F., 2008. Spectral characteristics of Hayabusa 2 near-Earth asteroid targets 162173 1999 JU3 and 2001 QC34. AJ 135, 1101-1105.

\bibitem[Vilas(2012)]{vilas12} Vilas, F., 2012. New spectral reflectance observations of Hayabusa 2 near-Earth asteroid target 162173 1999 JU3. American Astronomical Society, 44th DPS meeting, Abstract 102.03.

\end{thebibliography}



\clearpage	


\begin{landscape}
\begin{center}
\begin{singlespacing}
\begin{longtable}{llcccrccc}
\hline 
\hline
UT Date		& Instrument and Facility	& V		& $\Delta$ (AU)	& $R$ (AU)	& $\alpha$ ($^\circ$)		& \leftmoon	& T-O-M ($^\circ$) 	& Solar Analogs \\
\hline
1999-05-17$^\dagger$ & DSPEC Palomar 200" & 17.7	& 0.305		& 1.314		& 6.1					& 5\%		& 149			&  SA102-1081, 16 Cyg B \\
2007-07-11$^\ddagger$ & RCHAN MMT 6.5m & 20.3	& 0.569		& 1.381		& 40.3				& 12\%		& 78 				&  SA114-654 \\
2007-09-10$^\ddagger$ & RCHAN MMT 6.5m & 17.9	 & 0.273		& 1.254		& 22.5				& 2\%		& 150			&  SA114-654 \\
2007-09-11$^\ddagger$ & RCHAN MMT 6.5m & 17.9	 & 0.271		& 1.251		& 22.8				& 0\%		& 152			&  SA113-276, SA114-654 \\
2007-09-18	& SpeX IRTF			& 17.9	 & 0.260			& 1.231		& 26.5				& 37\%		& 90 				&  SA112-1333, SA115-271, SA93-101, Hya64 \\
2007-09-20	& SpeX IRTF			& 17.9	 & 0.258			& 1.226		& 27.9				& 56\%		& 68 				& SA110-361, SA112-1333, SA115-271, SA93-101 \\
2012-06-01	& LDSS3 Magellan 6.5m	& 17.7	 & 0.353			& 1.367		& 0.2					& 85\%		& 45				&  SA110-361, HD149182 \\
2012-06-02	& LDSS3 Magellan 6.5m	& 17.8	 & 0.354			& 1.368		& 1.0					& 92\%		& 31				&  HD149182 \\
2012-06-03	& LDSS3 Magellan 6.5m	& 17.9	 & 0.356			& 1.370		& 2.0					& 98\%		& 15				&  SA107-998 \\
2012-06-05	& FIRE Magellan 6.5m	& 18.1	 & 0.360			& 1.373		& 4.4					& 99\%		& 15				&  SA107-998 \\
2012-07-10	& FIRE Magellan 6.5m	& 19.9	 & 0.518			& 1.410		& 33.1				& 60\%		& 127			&  SA113-276 \\
\hline
\hline
\caption[]{Chronological summary of spectroscopic observations of 1999 JU3. \\
The columns in this table are the UT date of observation, the instrument and facility that were used, the apparent V-band magnitude (from the Minor Planet Center), geocentric distance ($\Delta$), heliocentric distance ($R$), the solar phase angle ($\alpha$), the phase of the moon (\leftmoon), the target-observer-Moon angle (T-O-M), and the solar analog stars used for calibration. The first four entries are from $^\dagger$\citet{binzel01} or $^\ddagger$\citet{vilas08}.
}
\label{tab.spec}
\label{lasttable}
\end{longtable}
\end{singlespacing}
\end{center}
\end{landscape}

\clearpage


\begin{center}
\begin{singlespacing}
\begin{longtable}{llcccr}
\hline 
\hline
UT Date		& Instrument and Facility	& V		& $\Delta$ (AU)	& $R$ (AU)	& $\alpha$ ($^\circ$)	\\
\hline
2011-10-30	& EEV 1k CCD Lulin 1m 	& 20.1	 & 0.438			& 1.002		& 76.1			  \\
2012-04-05	& IMACS Magellan 6.5m	& 19.8	 & 0.461			& 1.237		& 49.4			  \\
2012-04-06	& IMACS Magellan 6.5m	& 19.8	 & 0.457			& 1.240		& 48.9			  \\
2012-04-07	& IMACS Magellan 6.5m	& 19.8	 & 0.454			& 1.243		& 48.5			  \\
2012-05-31	& SITe1k Tenagra 0.8m	& 17.7	 & 0.352			& 1.366		& 0.7				  \\
2012-06-07	& SITe1k Tenagra 0.8m	& 18.2	 & 0.366			& 1.377		& 7.1				  \\
2012-06-08	& SITe1k Tenagra 0.8m	& 18.3	 & 0.368			& 1.378		& 8.2				  \\
2012-06-09	& SITe1k Tenagra 0.8m	& 18.4	 & 0.371			& 1.380		& 9.2				  \\
2012-06-20	& Tek1k Bosque Alegre 1.5m	& 19.0	 & 0.409		& 1.393		& 19.4			  \\
2012-06-21	& Tek1k Bosque Alegre 1.5m	& 19.0	 & 0.413		& 1.394		& 20.3			  \\
\hline
\hline
\caption[]{Summary of 1999 JU3 photometric observations. \\
The columns in this table are the same as in Table \ref{tab.phot}.
}
\label{tab.phot}
\label{lasttable}
\end{longtable}
\end{singlespacing}
\end{center}

\clearpage


\begin{center}
\begin{singlespacing}
\begin{longtable}{lccccrc}
\hline 
\hline
UT Midpoint		& Reference			& V		& $\Delta$ (AU)	& $R$ (AU)	& $\alpha$ ($^\circ$)	& Sub-Observer Long. and Lat. \\\hline
\hline
\hline
\multicolumn{4}{c}{\bf \citet{muller11} Shape Model}\\
1999-05-17 07:49	& B01				& 17.7	& 0.305			& 1.314		& 6.1				& $-169^\circ$, $19^\circ$ \\
2007-07-11 10:05	& V08				& 20.3	& 0.569			& 1.381		& 40.3				& $-40^\circ$, $8^\circ$ \\
2007-09-10 07:00	& V08				& 17.9	 & 0.273			& 1.254		& 22.5				& $169^\circ$, $25^\circ$ \\
2007-09-11 05:50	& V08				& 17.9	 & 0.271			& 1.251		& 22.8				& $172^\circ$, $25^\circ$ \\
2012-06-01 07:07	& A					& 17.7	 & 0.353			& 1.367		& 0.2				& $158^\circ$, $27^\circ$ \\
2012-06-01 08:09	& B					& 17.7	 & 0.353			& 1.367		& 0.2				& $109^\circ$, $28^\circ$ \\
2012-06-02 03:54	& C					& 17.8	 & 0.354			& 1.368		& 1.0				& $-102^\circ$, $28^\circ$ \\
2012-06-03 01:54	& D					& 17.9	 & 0.356			& 1.370		& 2.0				& $-59^\circ$, $28^\circ$ \\
2012-06-03 03:34	& E					& 17.9	 & 0.356			& 1.370		& 2.1				& $-137^\circ$, $28^\circ$ \\
2012-06-03 04:54	& F					& 17.9	 & 0.356			& 1.370		& 2.2				& $159^\circ$, $28^\circ$ \\
\multicolumn{4}{c}{\bf \citet{kawakami10} Shape Model}\\
1999-05-17 07:49	& B01				& 17.7	& 0.305			& 1.314		& 6.1				& $-19^\circ$, $10^\circ$ \\
2007-07-11 10:05	& V08				& 20.3	& 0.569			& 1.381		& 40.3				& $25^\circ$, $-71^\circ$ \\
2007-09-10 07:00	& V08				& 17.9	 & 0.273			& 1.254		& 22.5				& $-61^\circ$, $-77^\circ$ \\
2007-09-11 05:50	& V08				& 17.9	 & 0.271			& 1.251		& 22.8				& $-57^\circ$, $-78^\circ$ \\
2012-06-01 07:07	& A					& 17.7	 & 0.353			& 1.367		& 0.2				& $-34^\circ$, $-9^\circ$ \\
2012-06-01 08:09	& B					& 17.7	 & 0.353			& 1.367		& 0.2				& $-83^\circ$, $-9^\circ$ \\
2012-06-02 03:54	& C					& 17.8	 & 0.354			& 1.368		& 1.0				& $65^\circ$, $-9^\circ$ \\
2012-06-03 01:54	& D					& 17.9	 & 0.356			& 1.370		& 2.0				& $107^\circ$, $-8^\circ$ \\
2012-06-03 03:34	& E					& 17.9	 & 0.356			& 1.370		& 2.1				& $29^\circ$, $-8^\circ$ \\
2012-06-03 04:54	& F					& 17.9	 & 0.356			& 1.370		& 2.2				& $-35^\circ$, $-8^\circ$ \\
\hline
\caption[]{Sub-observer points for all published visible-wavelength spectra of 1999 JU3. \\
Columns in this table are the UT midpoint of observation, reference [B01 = \citet{binzel01}; V08 = \citet{vilas08}; letters A-F indicate the  corresponding spectra in Figure \ref{fig.spec}], solar phase angle ($\alpha$), and the sub-observer longitude and latitude. These data are presented for both shape models of 1999 JU3.
}
\label{tab.subobs}
\label{lasttable}
\end{longtable}
\end{singlespacing}
\end{center}


\begin{figure}[!t]
\begin{center}
\includegraphics[width=12cm]{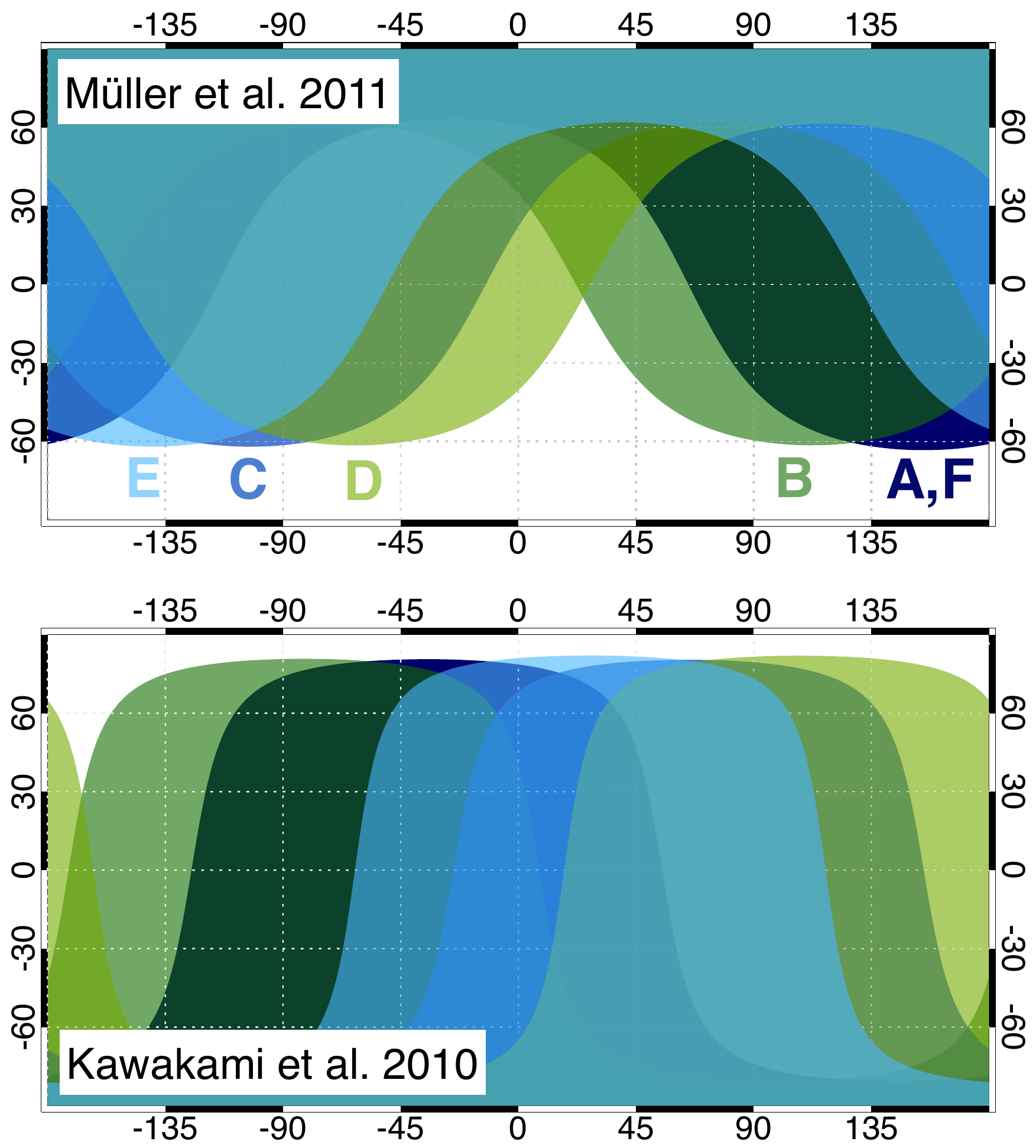}
\end{center}
\caption[]{SUPPLEMENTARY ONLINE FIGURE: Cylindrical projection of 1999 JU3's surface with sub-observer hemispherical footprints indicating the regions accessed by our spectra A-F. Projections are shown for both the M\"uller et al. (2011) and Kawakami et al. (2010) shape models. The hemispherical regions are color coded for each spectrum as indicated by the letters in the top panel. The sub-observer hemispheres for spectra A and F are so similar that they have been combined into a single dark blue region. The sub-observer points in Figure 5 would lie at the center of their respective regions. The Vilas (2008) and Binzel et al. (2001a) hemispheres are not plotted, but it is clear that they would almost entirely overlap with the regions here. Our spectra obtained hemispherical averages over a majority of the asteroid's surface, independent of which shape model is assumed, thus leaving little room for heterogenous regions that could produce prominent spectral absorption bands.
} 
\label{fig:without}
\label{lastfig}
\end{figure}

\end{document}